\documentstyle[aps,preprint,epsf,floats]{revtex}

\tighten  

\draft

\def\beq{\begin{equation}}
\def\eeq{\end{equation}}
\def\beqn{\begin{eqnarray}}
\def\eeqn{\end{eqnarray}}
\def\nn{\nonumber\\ }
\def\Re{\mbox{\,Re\,}}
\def\Im{\mbox{\,Im\,}}

\def\See{\vec\epsilon^{\prime *}\! \cdot\vec\epsilon\,}
\def\Sss{\vec s^{\prime *}\! \cdot\vec s\,}
\def\Ses{\vec\epsilon^{\prime *}\! \cdot\vec s\,}
\def\Sse{\vec s^{\prime *}\! \cdot\vec\epsilon\,}
\def\Vee{\vec\sigma\cdot\vec\epsilon^{\prime *}\! \times\vec\epsilon\,}
\def\Vss{\vec\sigma\cdot\vec s^{\prime *}\! \times\vec s\,}
\def\Ske{\vec\sigma\cdot\vec k \,}
\def\Sks{\vec\sigma\cdot\vec k'\,}

\def\vec#1{\mbox{\boldmath $#1$}}

\begin{document}

\preprint{\hfill\parbox[b]{0.3\hsize}
 {MKPH-T-97-16 \\ 
FTUV/97-41; IFIC/97-41 \\Submitted to Phys. Rev. C }}

\title{Polarization Phenomena in Small-Angle
Photoproduction of \\ $\bbox{e^+e^-}$ Pairs 
and the Gerasimov--Drell--Hearn Sum Rule }

\author{A.I. L'vov}
\address{P.N. Lebedev Physical Institute, Moscow, 117924, Russia}

\author{S. Scopetta\thanks{
Present address: Departament de Fisica Te\`orica,
Universitat de Val\`encia, 46100 Burjassot (Val\`encia), Spain}}
\address{Department of Physics, University of Perugia, I-06100 Perugia, Italy
\\ Institut f\"ur Kernphysik, Universit\"at Mainz, D-55099 Mainz, Germany}

\author{D. Drechsel and S. Scherer}
\address{Institut f\"ur Kernphysik, Universit\"at Mainz, D-55099
 Mainz, Germany \\ ~ }

\date{July 1997}
\maketitle

\begin{abstract}
   Photoproduction of $e^+e^-$ pairs at small angles is investigated as a
tool to determine the functions $f_1$ and $f_2$ entering the
real-photon forward Compton scattering amplitude.
   The method is based on an interference of the Bethe--Heitler and the
virtual Compton scattering mechanisms, generating an azimuthal asymmetry
in the $e^+$ versus $e^-$ yield.
   The general case of a circularly polarized beam and a
longitudinally polarized target allows one to determine both the
real and imaginary parts of $f_1$ as well as $f_2$.
   The imaginary part of $f_2$ requires target polarization only.
   We calculate cross sections and asymmetries of the reaction
$p(\gamma,e^+e^-)p$, estimate corrections and backgrounds, and
propose suitable kinematical regions to perform the experiment.
   Our investigation shows that photoproduction of $e^+e^-$-pairs off the
proton and light nuclei may serve as a rather sensitive test of the validity
of the Gerasimov--Drell--Hearn sum rule.
\end{abstract}


\section{Introduction}

   In studies of spin-dependent structure functions of nucleons and nuclei
with real and virtual photons, the verification of the
Gerasimov--Drell--Hearn sum rule (GDH) \cite{gdh1,gdh2} is of special
interest \cite{dre95,bas97}.
   Apart from general principles as gauge and Lorentz
invariance, unitarity, analyticity, and crossing symmetry, this
sum rule is based on the assumption that the forward Compton scattering
becomes spin independent at high energies.
   Therefore, the GDH provides an excellent test of the spin dynamics
of the nucleon.

   In the lab frame, the amplitude for forward Compton scattering off a
spin-1/2 target is described by two even functions $f_{1,2}$ of the
initial-photon energy $\omega$,
\beq
\label{f-forward}
  f = \See\,f_1(\omega) + i\omega\,\Vee f_2(\omega),
\eeq
which, at $\omega=0$, are constrained by the low-energy theorems
\cite{let1,let2}:
\beq
\label{let}
  f_1(0) = -\frac{\alpha Z^2}{M}, \qquad
  f_2(0) = -\frac{\alpha \kappa^2}{2 M^2}.
\eeq
   Here, $M$, $eZ$, and $\kappa$ are the mass, electric charge, and anomalous
magnetic moment of the target, respectively, and
$\alpha=e^2/4\pi \simeq 1/137$.
   Within the framework of the Regge pole model \cite{Regge,fri69}, these
functions behave like
\beq
  f_1(\omega) \propto \omega^{\alpha_R(0)},  \quad
  f_2(\omega) \propto \omega^{\alpha_R(0)-1} \quad
   \mbox{for} \quad \omega\to\infty,
\eeq
where $\alpha_R(0)$ is the intercept of the leading $t$-channel Regge
exchange contributing to the amplitudes.
   With the usual assumption of $\alpha_R(0)\alt 1$, both $f_1$ and
$f_2$ satisfy {\em once}-subtracted dispersion relations.
   The optical theorem allows one to express the imaginary parts of these
amplitudes in terms of the total photoabsorption cross sections
$\sigma_{1/2}$ and $\sigma_{3/2}$, where the subscripts refer to total
helicities 1/2 and 3/2, respectively,
\beq
\label{unitarity}
 \Im f_1=\frac{\omega}{8\pi}(\sigma_{1/2}+\sigma_{3/2})
        =\frac{\omega}{4\pi}\sigma_{\rm tot}, \quad
 \Im f_2=\frac{1}{8\pi}(\sigma_{1/2}-\sigma_{3/2})
        \equiv \frac{1}{4\pi}\Delta\sigma.
\eeq
   Therefore, we may write dispersion relations for $f_1$ and $f_2$,
\beq
\label{DR-f1}
  f_1(\omega) = - \frac{\alpha Z^2}{M}
  + \frac{\omega^2}{2\pi^2} \int_{\omega_{\rm thr}}^\infty
  \frac{\sigma_{\rm tot}(\omega')}{{\omega'}^2-\omega^2-i0^+}\,
     d\omega',
\eeq
\beq
\label{DR-f2}
  f_2(\omega) = - \frac{\alpha\kappa^2}{2 M^2}
  + \frac{\omega^2}{2\pi^2} \int_{\omega_{\rm thr}}^\infty
  \frac{\Delta\sigma(\omega')}{{\omega'}^2-\omega^2-i0^+}\,
     \frac{d\omega'}{\omega'}.
\eeq
   The GDH,
\beq
\label{GDH}
  \int_{\omega_{\rm thr}}^\infty
  \Big(\sigma_{3/2}(\omega)-\sigma_{1/2}(\omega)\Big)
    \,\frac{d\omega}{\omega} = \frac{2\pi^2\alpha\kappa^2}{M^2},
\eeq
arises from Eq.\ (\ref{DR-f2}) under the stronger assumption that
$f_2(\omega) \to 0$ for $\omega\to\infty$, or alternatively
$\alpha_R(0) < 1$.
   Numerical investigations \cite{kar73,wor92,bur93,san95} of the integral
on the lhs of Eq.\ (\ref{GDH}) in the case of the nucleon reveal a different
behavior for the two isospin combinations of the proton and neutron
amplitudes,
\beq
  f_{1,2}^{I=0} =\frac12(f_{1,2}^p+f_{1,2}^n),  \qquad
  f_{1,2}^{I=1} =\frac12(f_{1,2}^p-f_{1,2}^n),
\eeq
which correspond to isoscalar and isovector exchanges in the
$t$-channel.
   The GDH for $I=0$ seems to work very successfully,
supporting the conjecture that the leading isoscalar Regge exchange
(Pomeron) is decoupled from the spin-dependent transitions.
   It is very surprising, however, that the GDH for $I=1$ seems to be
violated, despite the fact that all known isovector exchanges (such as the
$a_1$ and $a_2$ mesons) have an intercept $\alpha_R(0) \alt 0.5$ which
supports the assumption of $f_2^{I=1}\to 0$ for $\omega\to\infty$.

   As was emphasized in \cite{san95}, the scale of the
violation in the isovector sector is not small.
   Evaluating the GDH integral, Eq.\ (\ref{GDH}), with the VPI multipoles
for single-pion photoproduction  (solution SP97K \cite{arn96})
and with the model of Ref.\ \cite{lvo97} for double-pion photoproduction,
up to the energy of 1~GeV we find for the $I=1$ component
\beq
\label{GDH,I=1}
   38.8+8.0=46.8~\mbox{$\mu$b} \quad \mbox{(GDH $I=1$, $\omega\le 1$ GeV)},
\eeq
   which has to be compared with $-14.2$ $\mu$b of the rhs of
Eq.\ (\ref{GDH}).
   In Eq.\ (\ref{GDH,I=1}), the $1\pi$- and $2\pi$-contributions are given
separately.
   Analogously, we obtain for the $I=0$ component of the GDH
\beq
\label{GDH,I=0}
  164.6+52.9=217.5~\mbox{$\mu$b} \quad \mbox{(GDH $I=0$, $\omega\le 1$ GeV)},
\eeq
which matches the value 218.9 $\mu$b of the rhs of Eq.\ (\ref{GDH}).
   The large positive $2\pi$ contribution in Eq.\ (\ref{GDH,I=0})
is related with the process $\gamma N \to \pi \Delta$ which is predominantly
isovector and contributes little to Eq.\ (\ref{GDH,I=1}).
   Assuming that the difference between Eq.\ (\ref{GDH,I=1}) and the GDH
prediction is resolved by an excess of the neutron over the proton
cross section at higher energies, e.g., between 1 to 3 GeV,
we would obtain $(\sigma_{3/2}-\sigma_{1/2})^n -
(\sigma_{3/2}-\sigma_{1/2})^p \simeq 110 ~\mu$b.
   Such a huge difference would be a great surprise, because the
total cross section is $\sigma_{\rm tot}\simeq 120{-}150$ $\mu$b at
these energies.

   We can think of two options to explain the discrepancy.
\begin{enumerate}
\item The GDH for $I=1$ is not valid, and we have the first
experimentally established example that the high-energy behavior of a
photon scattering amplitude is determined by the so-called fixed pole ($J=1$)
in the complex plane of the angular momentum \cite{Regge} rather
than known Regge exchanges with smaller intercepts $\alpha_R(0)$.
   In particular, this would violate the assumption of Vector Meson Dominance
in high-energy Compton scattering since fixed poles are forbidden in vector
meson photoproduction \cite{Regge}.
   Some arguments against the fixed pole are discussed in \cite{bas97}.
\item The existing experimental information on the spin and
isospin dependence of photoabsorption is not accurate enough.
   An alternative experiment to clarify the nature of the discrepancy
would be to measure the amplitude $f_2(\omega)$, Eq.\ (\ref{DR-f2}).
\end{enumerate}

   A straightforward way to study $\Im f_2$ is provided by measuring the
photoproduction cross sections $\sigma_{1/2}$ and $\sigma_{3/2}$ over a
wide energy region, using $4\pi$ detectors for the final hadrons.
   An independent determination of $\Re f_2$ by photon scattering would
also be of interest since it constrains the high-energy behavior of
$\Delta\sigma$ via the dispersion relation, Eq.\ (\ref{DR-f2}).
   In the present paper we propose an alternative method to determine both
$\Re f_2$ and $\Im f_2$ by measuring the photoproduction of small-angle
$e^+e^-$ pairs which serve as a polarimeter of the virtual photon in the
subprocess
\beq
\label{subprocess}
  \gamma N \to \gamma^* N'.
\eeq
   Since the conversion factor of the virtual photon into the $e^+e^-$
pairs is small ($\sim \alpha/3\pi$), this contribution will be strongly
suppressed compared with a huge background of Bethe-Heitler (BH) pair
production.
   However, the interference between the Bethe-Heitler amplitude and the
amplitude of virtual Compton scattering (VCS) leads to a specific azimuthal
asymmetry between yields of positrons and electrons which could be studied
with polarized photons and targets.
   A similar method was already used in Ref.~\cite{alv73} to determine $\Re
f_1$ at an energy of 2.2 GeV using an unpolarized beam and an unpolarized
target.

\section{Bethe--Heitler and virtual Compton scattering amplitudes}

Considering the reaction
\beq
\label{reaction}
  \gamma N \to e^+ e^- N'
\eeq
in the lab frame, we denote the energy, momentum, and helicity of the
positron by $\varepsilon_1$, $\vec p_1$, $\frac12 h_1$. Similarly, we
use $\varepsilon_2$, $\vec p_2$, $\frac12 h_2$ for the kinematical
variables of the electron and $\omega$, $\vec k$, $h_\gamma$ for the
photon.  The nucleon spin projections on the direction of the photon
beam are denoted by $\frac12 h_N$ and  $\frac12 h_N'$, and the masses
of the nucleon and leptons by $M$ and $m$, respectively.

Aside from an overall azimuthal angle, the kinematics of $e^+e^-$
photoproduction is specified by four variables.  We choose two of them
to be the invariant mass $W$ of the pair and the invariant momentum
transfer $Q$,
\beq
  W^2 = {k'}^2, \quad k' \equiv p_1+p_2, \quad
  Q^2=-q^2, \quad q \equiv p_N'-p_N.
\eeq
   The beam direction $\hat{\vec k}$ and the vector $\vec k'$ define the
reaction plane.
   Furthermore, we introduce the fractions of the virtual-photon energy
$\omega'$, which are carried by the positron and electron, respectively,
\beq
  x_1 = \frac{\varepsilon_1}{\omega'}, \quad
  x_2 = \frac{\varepsilon_2}{\omega'} = 1-x_1, \quad
  \omega' = \varepsilon_1+\varepsilon_2,
\eeq
and denote the azimuthal angles of the two particles, with respect to
the total momentum $\vec k'$ of the pair and the reaction plane, by
$\phi_1$ and $\phi_2=\pi+\phi_1$ (see Fig.~1).
   The kinetic energy and momentum of the recoiling nucleon read
\beq
  q_0 = \frac{Q^2}{2M}, \quad q_z=q_0 + \frac{W^2+Q^2}{2\omega},
 \quad q^2_\perp = Q^2 + q_0^2 - q_z^2 \ge 0,
\eeq
where the $z$-axis is directed along the photon beam.
   The energy and momentum carried by the pair are $\omega'=\omega-q_0$ and
$\vec k' = \vec k -\vec q$, respectively.  The polar angles between
$\vec p_{1,2}$ and $\vec k'$ are given by
\beq
   2|\vec p_i|\, |\vec k'| \cos\theta_i = 2\varepsilon_i \omega' - W^2,
  \quad i=1,2,
\eeq
and the angle $\theta$ between the beam and the total momentum of the
pair follows from
\beq
 \tan \theta = \frac{q_\perp}{\omega - q_z}.
\eeq

The differential cross section of reaction (\ref{reaction}) reads
\beq
  d\sigma = \frac{1}{(4\pi)^3}\,\frac{\omega'}{|\vec k'|}\,
  \frac{dW^2\,dQ^2}{16M^2\omega^2}\,dx_1\,\frac{d\phi_1}{2\pi} \,|T|^2,
\eeq
where the appropriate sum or average over spin projections is implied.

To lowest order in the electromagnetic coupling $e>0$, the amplitude
$T=T_{\rm BH}+T_{\rm VCS}$ consists of the contributions shown in
Fig.~2. Typically, we consider the kinematical region of
\beq
\label{kin}
 \omega \simeq 1 \mbox{~GeV}, \quad
 W\simeq 50 \mbox{~MeV}, \quad Q\simeq 100 \mbox{~MeV},
 \quad x_1 \simeq x_2 \simeq 0.5\,,
\eeq
corresponding to small angles $p_{1\perp}/\varepsilon_1 \ll 1$ and
$p_{2\perp}/\varepsilon_2 \ll 1$ between the photon beam and the
momenta of the leptons. We also assume that $p_{1\perp},p_{2\perp} \gg
m$. In this region the angles $\theta_i$,
\beq
  \sin^2 \theta_i = \frac 1{{\vec k'}^2 \vec p_i^2}
  \Big[ \varepsilon_1\varepsilon_2 W^2 - m^2 {\vec k'}^2
  -\frac{W^4}4 \Big] \simeq \frac{W^2}{\omega^2}\Big(\frac1{x_i}-1\Big),
\eeq
are proportional to the invariant mass $W$, and the total opening angle
of the pair is equal to
\beq
 \theta_1 + \theta_2 \simeq \frac W\omega (x_1 x_2)^{-1/2}.
\eeq

The BH amplitude is
\beq
\label{TBHexact}
  T_{\rm BH} = -\frac{e^3 J^\mu}{Q^2} \, \bar u(p_2,h_2) \Big[
  \gamma_\mu \frac{\gamma\cdot(k-p_1)+m}{(k-p_1)^2-m^2}
  \gamma\cdot\epsilon +
  \gamma\cdot\epsilon
  \frac{\gamma\cdot(p_2-k)+m}{(p_2-k)^2-m^2} \gamma_\mu
  \Big] v(p_1,h_1),
\eeq
where $\epsilon_\mu$ is the photon polarization vector and $J_\mu$ is
the nucleon electromagnetic current,
\beq
\label{J}
  J_\mu =  \bar u'(h_N') \Big\{ \gamma_\mu F_1(t) +
  \frac 1{4M} [\gamma\cdot q, \gamma_\mu] F_2(t) \Big\} u(h_N),
\eeq
with $t=-Q^2$ and the form factors normalized to $F_1(0)=Z$ and
$F_2(0)=\kappa$.

In the case of particles with definite helicities we follow the phase
conventions of Jacob and Wick (see, e.g., \cite{jac59,har72}),
\beq
\label{epsilon}
  \epsilon=(0; \vec \epsilon) = \frac 1{\sqrt2} (0; -h_\gamma,-i,0),
\eeq
and
\beq
\label{bi-spinors}
  u(p_2,h_2) = { \sqrt{\varepsilon_2+m} \choose h_2
  \sqrt{\varepsilon_2-m} }  \chi(p_2,h_2), \qquad
  v(p_1,h_1) = { -\sqrt{\varepsilon_1-m} \choose h_1
  \sqrt{\varepsilon_1+m} }  \chi(p_1,-h_1).
\eeq
Here $\chi(p,h)$ are Pauli spinors with helicity $\frac12 h$,
\beq
\label{spinors}
 \chi(p,+) = \frac 1{\sqrt{2p(p+p_z)}} {p+p_z \choose p_x + ip_y },
 \qquad
 \chi(p,-) = \frac 1{\sqrt{2p(p+p_z)}} {-p_x + ip_y \choose p+p_z }.
\eeq
For the nucleon polarized along the $z$-axis we have
\beq
\label{Nspinors}
  u(h_N) = \sqrt{2M} {1 \choose 0 } X(h_N), \qquad
  u'(h_N') = \frac 1{\sqrt{2M+q_0}}
  {2M+q_0 \choose \vec \sigma\cdot\vec q } X(h_N'),
\eeq
where
\beq
 X(+) = {1 \choose 0}, \qquad X(-) = {0 \choose 1}.
\eeq

   Using these formulae we compute the amplitude $T_{\rm BH}$ for the
appropriate polarizations of all 5 particles.
   For high energies and small angles of the produced leptons, we obtain a
good approximation by neglecting the lepton mass in Eqs.\ (\ref{TBHexact})
and (\ref{bi-spinors}) and the $q/M$ terms in Eqs.\ (\ref{J}) and
(\ref{Nspinors}), and replacing $p_z$ by $p$ in Eq.\ (\ref{spinors}).
   Keeping the leading terms in the transverse momenta, we arrive at the
simple formula
\beq
\label{TBH}
   T_{\rm BH} \simeq -\frac{4Me^3 F_1(t)}{Q^2}
  \omega\sqrt{x_1 x_2}\,  \, (x_1-x_2+h_1 h_\gamma) \,
  \vec\epsilon\cdot\vec D\,\delta_{h_1,-h_2}\,\delta_{h_N,h_N'}~,
\eeq
with
\beq
  \vec D = \frac{\vec p_{1\perp}}{p^2_{1\perp}}
         + \frac{\vec p_{2\perp}}{p^2_{2\perp}}
\eeq
and
\beqn
 && p^2_{1\perp} \simeq  x_1^2 Q^2
    + 2x_1\sqrt{x_1x_2}\,WQ\cos\phi_1 + x_1 x_2 W^2, \nn
 && p^2_{2\perp} \simeq  x_2^2 Q^2
    - 2x_2\sqrt{x_1x_2}\,WQ\cos\phi_1 + x_1 x_2 W^2 .
\label{p1p2-perp}
\eeqn
   In the kinematics of Eq.\ (\ref{kin}), Eq.~(\ref{TBH}) agrees
within 1$-$2\% with the full calculation of the largest matrix elements
($h_1=-h_2$, $h_N=h_N'$).
   The nucleon-spin-flip amplitudes which are not described by the
approximation (\ref{TBH}) are suppressed by a factor $Q/M$.
   They contribute incoherently to the differential cross section
summed over the final polarizations, and their contribution is only
${\simeq 2}\%$ at $Q=100$ MeV.
   The amplitudes with $h_1=h_2$ are suppressed by $m/p_{1,2\,\perp}$,
and their incoherent contribution is completely negligible except for the
region of very small $p_{1,2\,\perp}$ of about a few MeV which is beyond
the scope of our consideration.

   Next we consider the amplitude
\beq
   T_{\rm VCS} = \frac e{W^2} \epsilon^\mu T_{\mu\nu} l^\nu,
\eeq
where
\beq
\label{l}
 l_\nu = \bar u(p_2) \gamma_\nu v(p_1)
\eeq
is the lepton current.
   $T_{\mu\nu}$ is the amplitude of the subprocess (\ref{subprocess}), which
is generally described by 12 functions \cite{ber61} of the photon energy
$\omega$, the scattering angle $\theta$, and the virtual photon mass $W$.
   We assume, however, that $T_{\mu\nu}$ at small angles and photon masses
can be approximated by Eq.\ (\ref{f-forward}), i.e.\ by forward scattering
of real photons.
   Corrections to this approximation arise from (i) longitudinal
photon polarization, (ii) finite photon mass, and (iii) $\theta\ne 0$.
   We briefly discuss these effects in order.

(i) Using gauge invariance of the amplitude $M_\nu \equiv \epsilon^\mu
T_{\mu\nu}$, i.e.\ $M^\nu k'_\nu =0$, and current conservation, $l^\nu
k'_\nu =0$, we express the time-like components of $M_\nu$ and $l_\nu$
by the longitudinal ones (along $\vec k'$),
\beq
\label{M*l}
 M^\nu l_\nu \equiv M_0 l_0 -  M_L l_L - \vec M_T \cdot \vec l_T
 = - \frac {W^2}{{\omega'}^2} M_L l_L - \vec M_T \cdot \vec l_T .
\eeq
   Obviously, the longitudinal contribution vanishes for real photons,
$W^2 \to 0$.
   To get a more quantitative estimate, we note that the
space-like components of the lepton current, Eq.\  (\ref{l}), are given in the
small-angle approximation by
\beq
 l_L \simeq -2\omega\sqrt{x_1 x_2} \,\delta_{h_1,-h_2}
\eeq
and
\beq
\label{l_T}
 \vec l_T \simeq \sqrt{x_1 x_2} \,\delta_{h_1,-h_2}
 \Big[ (x_1-x_2) \vec d - i h_1 \hat {\vec k}\times \vec d\, \Big]~,
\eeq
where
\beq
  \vec d = \frac{\vec p_{1\perp}}{x_1}
         - \frac{\vec p_{2\perp}}{x_2}
\eeq
is directed perpendicular to the photon beam.
   (Note that the subscript $\perp$ denotes orthogonality to the $z$-axis,
the subscript $T$ orthogonality to the direction of $\vec k'$).
   Since the BH amplitude is essentially independent of the nucleon spin, see
Eq.~(\ref{TBH}), the interference of $T_{\rm BH}$ and $T_{\rm VCS}$ is
dominated by the no-spin-flip terms in $M_L$ and $\vec M_T$.
   Due to angular momentum conservation in the transition
$\gamma(h_\gamma=\pm1) N \to \gamma^*(h=0) N$, the no-spin-flip part of
$M_L$ has to vanish like $\theta$ for small angles.
   As a toy model, we may consider Compton scattering through E1-excitation
followed by the decay of a resonance with the quantum numbers of the
$D_{13}(1520)$.
   The corresponding amplitude
$\epsilon_\mu T^{\mu\nu} \epsilon^{\prime *}_\nu$ is proportional to
\beq
\label{D13model}
 \vec\epsilon^{\prime *}\cdot\vec S^+ \,\vec\epsilon\cdot\vec S =
 \frac 23 \See - \frac i3 \Vee = \frac{2+h_\gamma h_N}{3} \See,
\eeq
where $\vec S$ is the transition spin operator (see, e.g.\ \cite{ose94})
and $\vec\sigma$ is the nucleon spin operator, which can be replaced by
$h_N\hat{\vec k}$ for no-spin-flip transitions.
   Accordingly, in such transitions the amplitude
$\epsilon_i T_{ij} \epsilon_j^{\prime *}$ is
proportional to $\See$, and the L/T ratio reads
\beq
\label{L/T}
  \frac{|M_L|}{|\vec M_T|} = \frac 1{\sqrt2}\sin\theta
 \simeq \frac{Q}{\omega\sqrt2}  ~.
\eeq
   To be more accurate, we should do this calculation in the rest frame of
the resonance rather than in the lab frame.
   However, such a calculation does not change too much the estimate
(\ref{L/T}).

   Using the relationship
\beq
  \vec d^2 \simeq (x_1 x_2)^{-1} W^2,
\eeq
valid at high energies and small angles, we finally conclude that the
longitudinal contribution of $M_L$ to Eq.~(\ref{M*l}) is smaller than
the contribution of $M_T$ by a factor $\simeq WQ/\omega^2$, which is
typically $\alt 1\%$ in the kinematics of Eq.\ (\ref{kin}).

   It is worth noting that the longitudinal contribution in
Eq.~(\ref{M*l}) does not depend on the azimuthal direction of the
vector $\vec d$, and its interference with $T_{\rm BH}$ does not
contribute to azimuthal variations of positron and electron yields.
   The longitudinal contribution leads only to a small change of
the cross sections and asymmetries.

(ii) Another consequence of the virtuality of the photon $\gamma^*$ in
the subprocess (\ref{subprocess}) is the presence of additional form
factors at the photon-baryon vertices. We may expect that such form
factors would increase the amplitude $T_{\rm VCS}$ by a factor
$\simeq W^2/m_\rho^2$ which is $\alt 1\%$ for the kinematics considered.

(iii) In the general case of $\theta\ne 0$, the amplitude for real Compton
scattering is characterized by 6 independent functions
$A_i(\omega,\theta)$.
   Even for no-spin-flip transitions we still have
4 different amplitudes (see Appendix~A).
   Two of them describe no-photon-helicity-flip transitions and manifest
themselves in the amplitude $T_{\rm VCS}$ exactly like the functions $f_1$
and $f_2$ except for an additional angular dependence.
   The other two, $\tilde f_1$ and $\tilde f_2$, vanish at $\theta=0$ and
describe photon scattering involving a helicity flip.
   They result in a different azimuthal structure of
the BH--VCS interference, and give rise to $\cos 2\phi$ and $\sin 2\phi$
corrections to the formula discussed in the next section.

   The simple dipole model of Eq.\ (\ref{D13model}) suggests that the angular
dependence of $f_1$ and $f_2$ follows $1+\cos\theta$, such that the
$(\theta\ne 0)$-corrections are roughly of the scale $\theta^2 \simeq
1\%$.
   However, already at energies of 1 GeV many higher multipoles
contribute and the realistic angular dependence is steeper.
   In terms of the invariant momentum transfer $t=-Q^2$, the slope $B/2\simeq
3~\mbox{GeV}^{-2}$ of the amplitude $f_1(\omega,t)$ at high energies
should be close to the diffractive slope of the differential cross
section $d\sigma/dt \propto \exp(Bt)$ of Compton scattering
\cite{dud83}.
   Thus the decrease of $f_1$ at $Q\simeq 100$ MeV is ${\simeq} 3\%$.
   Nothing definite is known for $f_2(\omega,t)$ and for the flip amplitudes
$\tilde f_1$ and $\tilde f_2$.
   From dispersion calculations of the amplitudes $A_i$ \cite{lvo97}
we find that the amplitude $f_1$ at $\omega=1$ GeV decreases by 1.4\% and
3.9\% (real and imaginary parts, respectively), when $t$ varies from
0 to $-0.01~\mbox{GeV}^2$.
   For the amplitude $f_2$ the decrease is equal to 0.6\% and 5.2\%,
respectively.
   At $t=-0.01~\mbox{GeV}^2$ the helicity-flip amplitudes are very small,
$|\tilde f_1/f_1|=0.3\%$ and $|\tilde f_2/f_2|=0.15\%$.
   We conclude that even for a momentum transfer $Q$ of 100 MeV the Compton
scattering amplitude is not modified by more than ${\simeq}5\%$, at least at
$\omega=1$ GeV.

   Eventually, we evaluate the amplitude $T_{\rm VCS}$ by keeping only the
term $-\vec M_T \cdot \vec l_T$ in Eq.\ (\ref{M*l}) with
\beq
 -\vec M_T = 8\pi M \vec\epsilon \, ( f_1 -
  h_\gamma h_N \omega f_2 ) \,\delta_{h_N,h_N'} \, ,
\eeq
   where the factor $8\pi M$ provides the correct normalization and the term
with $f_2$ is treated in the no-spin-flip approximation $\vec\sigma \to
h_N\hat{\vec k}$.
   Using Eq.\ (\ref{l_T}), we finally obtain
\footnote{In the first report on our calculation \cite{VCS96}
the amplitude $T_{\rm VCS}$ was given a wrong sign.}

\beqn
\label{TVCS}
 && T_{\rm VCS} \simeq \frac{8\pi M e}{W^2}
  \sqrt{x_1 x_2}\, \vec\epsilon\cdot\vec d \, (x_1-x_2+h_1 h_\gamma) \,
  \nn && \hspace*{6em} ~ \times
  \Big(f_1(\omega) - h_\gamma h_N \omega f_2(\omega)\Big)
   \delta_{h_1,-h_2}\,\delta_{h_N,h_N'}.
\eeqn
{}From the above arguments we expect that this approximation to
$T_{\rm VCS}$ is sufficient to describe interference effects between
the BH and VCS amplitudes with a relative accuracy of a few per cent.

\section{Azimuthal asymmetries and determination of $f_2$}

   Generally, the VCS amplitude, Eq.\ (\ref{TVCS}), is small in comparison
with the BH amplitude, Eq.\ (\ref{TBH}), and can be seen only through
BH--VCS interference which, in turn, can be observed because of a specific
signature of odd $C$-parity of the leptons ($C_L$ parity).
   Under charge conjugation the lepton pair and the electromagnetic field
transform as
\beqn
 |e^+(p_1,h_1), e^-(p_2,h_2) \rangle &\stackrel{C}{\to}&
 |e^-(p_1,h_1), e^+(p_2,h_2) \rangle = -
 |e^+(p_2,h_2), e^-(p_1,h_1) \rangle ~,\nonumber\\
 A_\mu(x)&\stackrel{C}{\to}&-A_\mu(x).
\eeqn
   Thus the relevant amplitudes for a process where the lepton pair is
produced by a {\em single} virtual photon, e.g.\ in VCS or in the background
reaction
\beq
\label{pi0backgr}
 \gamma p \to \pi^0 X \to e^+e^-\gamma X,
\eeq
are even under the interchange
\beq
\label{C}
  \varepsilon_1,\vec p_1,h_1 \leftrightarrow \varepsilon_2,\vec p_2,h_2 \,.
\eeq
   On the other hand, the BH reaction involves the interaction of the
pair with {\em two} photons and, therefore, the BH
amplitude is odd under Eq.\ (\ref{C}).
   Thus its interference with the VCS contribution leading to the {\em same}
final state is antisymmetrical under the exchange $1 \leftrightarrow 2$.

   The BH cross section, summed over the polarizations of the
lepton pair and the recoil proton, has been calculated for various
kinematical situations (see, e.g., Fig.~3). It can be directly obtained
from the amplitude (\ref{TBH}). In order to check the numerical
calculations and to obtain simple analytical expressions, the following
approximations are useful.  We define, as an estimate for the yield,
\beq
\label{csBH}
  \sigma_{12}^{\rm BH} \equiv
  2\pi WQ \frac{d^4\sigma^{\rm BH}} {dW dQ\,dx_1 d\phi_1}
  \simeq \frac{8 \alpha^3 Z^2 W^2}{Q^2}
  \, x_1 x_2\,(x_1^2 + x_2^2) \vec D^2\,,
\eeq
which is symmetric under $1 \leftrightarrow 2$.  Note that
\beq
  \vec D^2 \simeq \frac{Q^2}{p^2_{1\perp} p^2_{2\perp}}~,
\eeq
where $p^2_{1,2\,\perp}$ are given in Eq.\ (\ref{p1p2-perp}).
   In the particular case of $x_1=x_2=1/2$,
\beq
  |\vec D(\phi_1=\pm\frac \pi 2)| \simeq  \frac{4Q}{Q^2+W^2},
\qquad
  |\vec D(\phi_1=0\mbox{~or~}\pi)| \simeq \frac{4Q}{|Q^2-W^2|}.
\eeq

   In the total cross section, the excess of positrons over electrons at
some angles is due to the $C_L$-odd interference term,
\beq
\label{12-21}
 \sigma_{12} - \sigma_{21}  \approx
  - \frac{16\alpha^2 Z}{\omega} \, x_1 x_2\,(x_1^2 + x_2^2)
  |\vec d|\,|\vec D|\, F(\omega,\phi),
\eeq
where the function
\begin{eqnarray}
  F(\omega,\phi) &=&
 (\Re f_1(\omega)-h_\gamma h_N \omega \Re f_2(\omega)) \cos\phi
  \nn && \qquad {} +
 (-h_\gamma \Im f_1(\omega)+ h_N \omega \Im f_2(\omega)) \sin\phi
\label{F}
\end{eqnarray}
depends on the azimuthal angle $\phi$ between the vectors $\vec d$ and
$\vec D$ and carries information on the real and imaginary parts of
$f_{1,2}$. The azimuth of $\vec d$ coincides with $\phi_1$, but the
vector $\vec D$ in general does not lie in the reaction plane,
\beq
   \phi = \phi_1 - \phi_D,\quad
   \sin\phi_D \simeq  - \frac{W\sin\phi_1}{p_{1\perp}p_{2\perp}}
   \Big(2x_1x_2 W\cos\phi_1 + (x_1-x_2)\sqrt{x_1x_2}\,Q\Big).
\eeq
   In particular, $\phi$ and $\phi_1$ are the same when the
pair is symmetric ($x_1=x_2$), and $\vec{p}_1$ is in the reaction
plane ($\phi_1=0$ or $\pi$, $Q>W$) or transversely to the plane
($\phi_1=\pm\pi/2$).

The function $F(\omega,\phi)$ can be measured through the $e^+$-$e^-$
asymmetry,
\beq
 \Sigma_{12} =
 \frac{\sigma_{12} - \sigma_{21}} {\sigma_{12} + \sigma_{21}}
 \simeq \frac{ 2\Re [T_{\rm VCS} T_{\rm BH}^*]} {|T_{\rm BH}|^2}
 \simeq -\frac{Q^2}{\alpha Z\omega W} \,
  \frac{F(\omega,\phi)}{\sqrt{x_1 x_2}\, |\vec D|},
\eeq
provided that background contributions such as (\ref{pi0backgr}) are
subtracted.
   The available knowledge of $f_1$ can be used to test this
procedure.

   Corresponding to the helicity structure of the function $F$, we may
introduce 4 different asymmetries $A$, $A_\gamma$, $A_N$, and
$A_{\gamma N}$,  such that
\beq
  \Sigma_{12} = A + h_\gamma A_\gamma + h_N A_N
     + h_\gamma h_N A_{\gamma N}.
\eeq
   The indices of these coefficients denote which particles are polarized.
   The asymmetries $A$ and $A_{\gamma N}$ are proportional to $\Re f_1$
and $\Re f_2$, respectively, and can be measured by observing the
leptons in the reaction plane.
   The asymmetries $A_\gamma$ and $A_N$ are proportional to
$\Im f_1$ and $\Im f_2$, respectively,
and can be measured by detecting leptons emitted transversely to the
reaction plane.

   The amplitudes $f_1$ and $f_2$ (see Fig. 4) and the asymmetries
(see Fig. 5) are estimated by using unitarity (\ref{unitarity}) and the
dispersion relations (\ref{DR-f1}) and (\ref{DR-f2}).
   The amplitudes $\Im f_1$ and $\Re f_1$ can be reliably calculated on
the basis of the total photoabsorption data for the proton
\cite{McC96,Arm72}, and the quantity $\Delta\sigma$ has been evaluated
by using the model of \cite{lvo97}, accounting for our present
knowledge of single-pion photoproduction multipoles \cite{arn96} and
resonance and background mechanisms of double-pion photoproduction.
   We note that these calculations are only performed in
order to illustrate the corrections due to $A_N$ and $A_{\gamma N}$.
    The asymmetries $A$ and
$A_\gamma$ should be measured and could be used for calibration
purposes.

   We emphasize that even with unpolarized photons one can
measure the GDH cross section $\Delta\sigma \propto \Im f_2$.
   If the photon hits the target polarized along the
beam direction {\em and} the
spin-3/2 photoabsorption cross section $\sigma_{3/2}$ is bigger than
$\sigma_{1/2}$, the positron is preferentially emitted to the rhs
of the reaction plane (i.e.\ $\vec k\times\vec k'\cdot\vec p_1 >0$).
   The corresponding asymmetry for $x_1=x_2$,
\beq
\label{A_N}
 A_N (\phi_1=\frac \pi 2) \simeq
 -\frac{Q (Q^2+W^2)}{8\pi \alpha Z W} \Delta\sigma,
\eeq
is sensitive to $\Delta\sigma$, with a proportionality factor independent of
the photon energy.
   In the kinematics of Eq.\ (\ref{kin}), a value $\Delta\sigma=\mp 29~\mu$b
leads to an asymmetry $A_N=\pm1\%$.
   Therefore it is necessary to measure $A_N$ with the accuracy
$\delta A_N\simeq 10^{-3}$ in order to obtain useful constraints
on $\Delta\sigma$.

   As can be seen from Eq.\ (\ref{DR-f2}), the convergence of the GDH
integral to its canonical value, Eq.\ (\ref{GDH}), and its saturation by
energies $\omega^2\ll \omega^2_0$, can happen if
and only if $f_2(\omega)\ll f_2(0)$ for $\omega\agt\omega_0$.
   Given, for example, $R\equiv \Re f_2(\omega) / f_2(0)=\pm 5\%$,
we expect to see the asymmetry
\beq
  A_{\gamma N}(\phi_1=0,\, x_1=x_2)= -R\,
  \frac{\kappa^2 Q}{4M^2 W}\,(Q^2-W^2) = \mp 0.07\%
\eeq
in the kinematics of Eq.\ (\ref{kin}) independently of $\omega$.
   Therefore, an experiment determining whether the asymmetries
$A_{\gamma N}$ and $A_N$ at energies of a few GeV are less than
${\simeq} 10^{-3}$ would provide a clean test of the validity of the
GDH sum rule at the accuracy level of a few per cent.

   Combining the helicity amplitudes (\ref{TBH}) and (\ref{TVCS})
with $h_\gamma=\pm 1$, we obtain cross sections and
asymmetries for linearly-polarized photons.
   Introducing $\vec s=\hat{\vec k}\times\vec\epsilon$, we obtain
cross sections by substituting, in Eq.~(\ref{csBH}),
\beq
  (x_1^2 + x_2^2) \vec D^2 \to
    (x_1-x_2)^2 (\vec\epsilon\cdot\vec D)^2 + (\vec s\cdot\vec D)^2,
\eeq
and the asymmetry by replacing, in Eq.~(\ref{12-21}),
\beqn
  && (x_1^2 + x_2^2) |\vec d|\,|\vec D| \,F(\omega,\phi) \to
  \Big[(x_1-x_2)^2 \,\vec\epsilon\cdot\vec d \,\vec\epsilon\cdot\vec D
    + \vec s\cdot\vec d \,\vec s\cdot\vec D \Big] \,\Re f_1 \nn
  && \qquad\qquad {} +
  \Big[(x_1-x_2)^2 \,\vec s\cdot\vec d \,\vec\epsilon\cdot\vec D
    - \vec\epsilon\cdot\vec d \,\vec s\cdot\vec D)\Big] \,h_N\omega\Im f_2~.
\eeqn
   As in the case of unpolarized photons, only $\Re f_1$ and
$\Im f_2$ can be measured in this way.

   For the sake of completeness we mention that the same interference
structures exist in the crossed reaction
\beq
\label{crossed}
  e^-p\to e^-p\gamma,
\eeq
which recently attracted considerable interest as a tool to study
VCS \cite{vcsproc}.
   The corresponding BH and VCS amplitudes in the
small-angle approximation are obtained similarly to Eqs.\ (\ref{TBH}) and
(\ref{TVCS}), or via the crossing symmetry
($\varepsilon_1,\vec p_1,h_1$ $\to$ $-\varepsilon,-\vec p,-h$,~
 $\varepsilon_2,\vec p_2,h_2$ $\to$ $ \varepsilon',\vec p',h'$, and
 $\omega,\vec\epsilon,h_\gamma$ $\to$ $-\omega,-\vec\epsilon^*,-h_\gamma$;
 the sign of minus in $\vec\epsilon^*$ is related to the definition
 of the helical vectors).
They are
\beq
 T_{\rm BH} \simeq \frac{4Me^3 F_1(t)}{-t}\,
  \frac{\sqrt{\varepsilon\varepsilon'}}{\omega}
  (\varepsilon+\varepsilon'+\omega h_\gamma h)
  \Big(-\frac{\vec\epsilon^*\cdot\vec p}{p_\perp^2}
       +\frac{\vec\epsilon^*\cdot\vec p'}{p_\perp^{\prime 2}}
  \Big) \,
  \delta_{h,h'}\,\delta_{h_N,h_N'}~,
\eeq
where the transverse components and the nucleon spin projections are
considered with respect to the direction of the final photon momentum
$\vec k$, and
\beqn
 && T_{\rm VCS} \simeq \frac{8\pi Me}{-(p-p')^2}\,
  \frac{\sqrt{\varepsilon\varepsilon'}}{\omega}
  (\varepsilon+\varepsilon'+\omega h_\gamma h)
  \Big( \frac{\vec\epsilon^*\cdot\vec p}{\varepsilon}
       -\frac{\vec\epsilon^*\cdot\vec p'}{\varepsilon'} \Big)
  \nn && \qquad\qquad ~\times
  (f_1(\omega) - h_\gamma h_N\omega f_2(\omega)) \,
  \delta_{h,h'}\,\delta_{h_N,h_N'}~.
\eeqn
   However, the important difference is that there is no natural
way to measure the $C_L$-odd effects in the reaction (\ref{crossed}),
except by using also a positron beam.

\section{Yields and backgrounds}

   With regard to an optimal choice of $W$ and $Q$, we note that the
yield estimate of the lepton pairs (cf.~(\ref{csBH})) is proportional to
$W^2\vec D^2/Q^2$.
   The statistical fluctuations in the asymmetry,
$\delta\Sigma_{12}\propto Q/W|\vec D|$, are by a factor of
$Q$ smaller than the asymmetry, $\Sigma_{12}\propto Q^2/W|\vec D|$.
   Thus it is profitable to choose $Q$ as large as possible, keeping $Q$ still
small enough to fit the condition imposed by a small scattering angle
$\theta$ as discussed in the previous section.
   The choice of $W$ is less important, provided that $W$ is still close
enough to the real-photon point.
   On the other hand, $W$ should not be less then ${\simeq}10$ MeV, since
otherwise multiple Coulomb scattering of the leptons in the target would
prevent an accurate measurement of small transverse momenta.

   In the kinematics of Eq.\ (\ref{kin}), which satisfies all of the above
requirements,  the BH cross section (\ref{csBH}) is a few tens of a nanobarn
(see Fig.~3).
   With an untagged photon beam intensity of
$I_\gamma \simeq 10^9~\mbox{s}^{-1},$ and a target of density
$N \simeq 5\cdot 10^{23}~\mbox{cm}^{-2}$,
the count rate is expected to reach a few events per second.
   This should make it possible to measure asymmetries with an
accuracy ${\simeq} 10^{-3}$.

   Most of the BH pairs are produced at very forward angles, and the
experimental setup should be insensitive to them.
   The average load of a single lepton detector is determined by the
inclusive BH cross section integrated over the directions of the second lepton
\cite{glu53},
\beq
\label{csBH-single-arm}
  2\pi p_{1\perp} \frac{d^3\sigma(\gamma p\to e^+X)}
      {dp_{1\perp}dx_1 d\phi_1}
  \simeq  \frac{8Z^2\alpha^3}{p_{1\perp}^2}
  \Big[(x_1^2+x_2^2)\Big(\log\frac{2x_1x_2\omega}{m}-\frac12\Big)
    +x_1x_2 \Big] .
\eeq
   In the kinematics of Eq.\ (\ref{kin}) with $p_{1\perp}\simeq Q/2
\simeq 50$ MeV, the rhs of Eq.\ (\ref{csBH-single-arm}) is equal to
1.7 $\mu$b, leading to a counting rate of $10^3~\mbox{s}^{-1}$.

   The reaction (\ref{pi0backgr}) generates azimuthally symmetric pairs
which dilute all the asymmetries.
   However, the distribution is less peaked at small total transverse
momentum $q_\perp = |\vec p_{1\perp} +\vec p_{2\perp}|$ than in the case
of BH pairs.
   The decay of relativistic pions in flight, $\pi^0\to e^+e^-\gamma$,
yields pairs distributed in the invariant mass $W$, the total energy
$\omega'$, and the relative energies $x_i=\varepsilon_i/\omega'$
according to
\beq
\label{pi0decay}
  \frac{d\Gamma(\pi^0\to e^+e^-\gamma)}{\Gamma(\pi^0\to \gamma\gamma)}
  \simeq  \frac{\alpha}{\pi}\,(1-\eta)^2\,\frac{dW^2}{W^2}
   \, \frac{d\omega'}{\varepsilon_\pi}\,(x_1^2+x_2^2)\,dx_1 ~,
\eeq
where $\varepsilon_\pi$ is the pion energy, $\eta=W^2/m_\pi^2$,
and $\eta\varepsilon_\pi<\omega'<\varepsilon_\pi$.
   Since the internal transverse momentum of the pair,
$ p_{\rm int} = (m_\pi/\varepsilon_\pi)
\sqrt{(\omega'-\eta\varepsilon_\pi) (\varepsilon_\pi-\omega')}
 \le (1-\eta)m_\pi/2$,
is smaller than the pion momentum, we neglect the effect of $ p_{\rm int}$
on the angular distribution of the pairs.
   Folding the differential cross section
$d^2\sigma(\omega,\varepsilon_\pi,\theta)/d\varepsilon_\pi d\Omega_{\rm
lab}$ of inclusive $\pi^0$ photoproduction with the photon beam
spectrum $I_\gamma d\omega/\omega$ and the decay distribution
(\ref{pi0decay}), we find the cross section $\sigma_{\rm backgr}
d\omega' / \omega'$ for producing a pair of energy $\omega'$ and
central angle $\theta\simeq Q/\omega'$ by photons with energies
between $\omega'$ and $\omega_{\rm max}$,
\beqn
  &&  \sigma_{\rm backgr} \equiv
  2\pi WQ \frac{d^4\sigma_{\rm back}}{dW dQ\,dx_1 d\phi_1}
   \simeq  4\alpha\omega'\theta^2 \,(1-\eta)^2 \, (x_1^2 + x_2^2) \,
   \nn && \qquad\qquad ~ \times
   \int_{\omega'}^{\omega_{\rm max}}  \frac{d\omega}{\omega}
   \int_{\omega'}^{\min(\omega'/\eta,~\omega)}
  \frac{d^2\sigma(\omega,\varepsilon_\pi,\theta)}
    {d\varepsilon_\pi d\Omega_{\rm lab}}
   \,\frac{d\varepsilon_\pi}{\varepsilon_\pi} ~.\quad~
\eeqn
As a numerical example, we consider the case of $\omega'=600$ MeV,
$W=50$ MeV, $Q=100$ MeV, $x_1=x_2=0.5$, and $\omega_{\rm max}=900$ MeV.
Using the unpolarized differential cross section $d\sigma/d\Omega_{\rm
lab} \simeq 2\mbox{~$\mu$b$/$sr}$ of the exclusive reaction $\gamma
p\to\pi^0 p$ at $\theta\simeq 9.5^\circ$ as a substitute for the
inclusive one, we obtain $\sigma_{\rm backgr} = 0.20$ nb.

\section{Summary}

   Recently, virtual Compton scattering off the nucleon has attracted
considerable interest (for an overview see \cite{vcsproc}).
   At low energies,  the so-called generalized polarizabilities
\cite{gui95} are accessible in the reaction $e^-p\to e^-p\gamma$,
whereas investigations at high energies allow one to study the wave
function of the valence quarks in the nucleon \cite{kroll96}.
   On the other hand, photoproduction of lepton pairs in the reaction
$\gamma p\to pe^+ e^-$ has been proposed to test vector meson dominance
for the nucleon \cite{schae95}, to obtain information on
the time-like form factors of the nucleon \cite{schae95,diep97,kor97},
and to investigate the longitudinal response of resonances \cite{kor97}.

   In this work we have suggested to study pair production for yet
another reason.
   We have argued that small-angle photoproduction of $e^+e^-$ pairs
may be used to measure the forward scattering amplitude of real
photons.
   The mechanism which makes such investigations possible is an
interference between the Bethe-Heitler amplitude and the virtual
Compton scattering amplitude leading to an asymmetry between the
electron and positron angular distributions.
   A determination of the real and imaginary parts of $f_1$ and $f_2$
requires a longitudinally polarized target and circularly polarized photons.
   The imaginary part of $f_2$, which is
of particular importance for testing the validity of the
Gerasimov-Drell-Hearn sum rule, can be measured with target
polarization only.

   We have proposed specific kinematical conditions which would
be particularly appropriate for a determination of $f_2$.
   The uncertainties due to a longitudinal photon
polarization, finite photon mass, and the emission of the virtual
photon at small non-forward angles have been estimated.
   We have argued that for the kinematics proposed the corresponding
errors are small.
   Our estimate has also shown that possible backgrounds, essentially from
$\pi^0$ decays, give rise to small effects only.

   In conclusion, photoproduction of $e^+e^-$ pairs provides a possibility
to determine the real-photon forward scattering amplitude.
   In particular, this process may be regarded as an alternative to
proposed experiments to test the Gerasimov-Drell-Hearn sum rule by
cross section measurements.

\acknowledgments
A.L. thanks the theory group of the Institut f\"ur Kernphysik and the
SFB 201 for their hospitality and support during his stay in Mainz
where a part of this work was done.
S.S. thanks the TMR programme of the European Commision ERB FMRX-CT96-008
for partial financial support.

\appendix
\section{Real Compton scattering at non-forward angles}

In terms of the invariant amplitudes $A_i(\omega,t)$ \cite{lvo97} the
amplitude for real Compton scattering has the following form in the lab
frame:
\beqn
T_{\rm CS}=\frac{\omega\omega'}{N(t)} && \, \Big\{
      2M\See
    \Big[ N^2(t)({-A_1}-A_3)-\frac{\nu^2}{M^2}A_5 - A_6\Big] \nn
   &&~+ 2M\Sss
    \Big[ N^2(t)(A_1-A_3)+\frac{\nu^2}{M^2}A_5 - A_6\Big] \nn
   &&~- 2i\nu\Vee (A_5+A_6) ~+~ 2i\nu\Vss (A_5-A_6) \nn
   &&~+ i \Ske\Sse
    \Big[ A_2+(1-\frac{\omega'}{M})A_4+\frac{\nu}M A_5 + A_6\Big] \nn
   &&~- i \Sks\Ses
    \Big[ A_2+(1+\frac{\omega }{M})A_4-\frac{\nu}M A_5 + A_6\Big] \nn
   &&~- i \Ske\Ses
    \Big[ {-A_2}+(1-\frac{\omega'}{M})A_4-\frac{\nu}M A_5 + A_6\Big] \nn
   &&~+ i \Sks\Sse
    \Big[ {-A_2}+(1+\frac{\omega }{M})A_4+\frac{\nu}M A_5 + A_6\Big]
      \Big\}.
\label{TCS}
\eeqn
   Here $\omega'=\omega/(1+(1-z)\omega/M)$ is the final photon energy,
$z=\cos\theta$, $\nu=(\omega+\omega')/2$, $t=-2\omega\omega'(1-z)$ is
the invariant momentum transfer squared, and $N(t)=(1-t/4M^2)^{1/2}$.
   The two magnetic vectors $\vec s$ and $\vec s'$ are defined as
\beq
 \vec s =\hat{\vec k} \times\vec\epsilon = -ih_\gamma\vec\epsilon,
 \quad
 \vec s'=\hat{\vec k'}\times\vec\epsilon'= -ih_\gamma'\vec\epsilon',
\label{s}
\eeq
for helical vectors $\vec\epsilon$ and $\vec\epsilon'$ as defined in
Eq.~(\ref{epsilon}), with $\vec\epsilon^{\prime *}\cdot\vec\epsilon =
(1+h_\gamma h_\gamma' z)/2$.

   Since we are interested in no-spin-flip transitions which interfere
with the essentially no-spin-flip Bethe--Heitler amplitude $T_{\rm BH}$,
we set $\vec\sigma \to h_N\hat{\vec k}$ in the above equation.
   We have to distinguish two cases:

i) No photon helicity flip ($h_\gamma'=h_\gamma$).  In this case the amplitude
(\ref{TCS}) becomes
\beq
 T_{\rm CS} \equiv 8\pi M \Big(f_1(\omega,t)
       - h_\gamma h_N \omega f_2(\omega,t)\Big),
\eeq
where
\beqn
  f_1(\omega,t) &=& -\frac{(1+z)\omega\omega'}{4\pi N(t)}
   \,\Big(A_6(\omega,t)+A_3(\omega,t) N^2(t)\Big),
 \nn
  f_2(\omega,t) &=& \frac{(1+z){\omega'}^2}{8\pi M\,N(t)}
    \,\Big((z-1)A_6(\omega,t)+(z+1)A_4(\omega,t)\Big).
\label{f1f2}
\eeqn
At energies $\omega$ of a few GeV and small $t$ we may expect that the
slope $B/2$ of the amplitude $f_1(\omega,t)$ considered as a function
of $t$ follows the diffractive behavior of the differential cross
section of Compton scattering, $d\sigma/dt \propto \exp(Bt)$ with
$B\simeq 6$ GeV$^{-2}$ \cite{dud83}.  Meanwhile, the high energy
behavior of the amplitude $f_2$, including its slope, is unknown.

ii) Photon helicity flip ($h_\gamma'=-h_\gamma$).  In this case the
amplitude vanishes at $t=0$, and the leading term
$O(t)$ is
\beq
 T_{\rm CS} \equiv 8\pi M \Big(\tilde f_1(\omega,t)
       - h_\gamma h_N \omega \tilde f_2(\omega,t)\Big),
\eeq
where
\beqn
  \tilde f_1(\omega,t) &\simeq& \frac{t}{8\pi}
   \,\Big(A_1(\omega,0)+\frac{\omega^2}{M^2}\,A_5(\omega,0)\Big),
   \nn
  \tilde f_2(\omega,t) &\simeq& \frac{t}{8\pi M}
  \,\Big(1+\frac{\omega}{M}\Big)\,A_5(\omega,0).
\label{f1f2tilde}
\eeqn
   The high-energy behavior of the helicity-flip amplitudes is also unknown.
   As discussed in \cite{lvo97}, data on electromagnetic polarizabilities of
the nucleon put some constraints on the amplitude $A_1(\omega,0)$.

   In the energy region below ${\simeq} 1$ GeV, the amplitudes $A_i$ are
dominated by nucleon resonances, other known channels of the nucleon
photoexcitation, and a few $t$-channel exchanges (mostly $\pi^0$ and
$\sigma$).
   Using unitarity, dispersion relations, and plausible assumptions on the
high-energy behavior of the amplitudes $A_1$ and $A_2$,
one can determine all the $A_i$ and describe the available data on
Compton scattering with {\em unpolarized} and {\em single}-polarized
particles \cite{lvo97}.
   Using these results as they are, we find all
the amplitudes $f_1$, $f_2$, $\tilde f_1$, and $\tilde f_2$ through
(\ref{f1f2}), (\ref{f1f2tilde}) and directly estimate their
$t$-dependence.

\newpage
\appendixonfalse
\section*{Figure Captions}

\vspace{1em}\noindent
{\bf Fig. 1}:
Kinematics of the reaction $\gamma N \rightarrow e^+ e^- N'$.  The
kinematical variables are described in the text.

\vspace{1em}\noindent
{\bf Fig. 2}:
Diagrams contributing to lepton pair production: Bethe--Heitler
mechanism (left) and virtual Compton scattering (right).

\vspace{1em}\noindent
{\bf Fig. 3}:
The Bethe--Heitler cross section (\ref{csBH}) for the proton as a
function of the invariant mass $W$ of the pair at incident photon energy
$\omega$=1000 MeV, azimuthal angle $\phi_1=90^\circ$, $x_1=x_2=0.5$, and for
three values of 4-momentum transfer, $Q = 75$ MeV (dashed curve),
$Q=100$ MeV (full curve), and $Q=125$ MeV (dotted curve).

\vspace{1em}\noindent
{\bf Fig. 4}:
The real and imaginary parts of the no-spin-flip and spin-flip
amplitudes $f_1$ and $f_2$, respectively, as functions of the beam
energy $\omega$. The calculations are described in the text.

\vspace{1em}\noindent
{\bf Fig. 5}:
The four asymmetries $A_\gamma$, $A_N$, $A$  and $A_{\gamma N}$ as
functions of the beam energy $\omega$, at fixed value of the invariant mass
of the pair, $W=50$ MeV, and at $x_1=x_2=0.5$. Dashed curve: $Q=75$
MeV, full curve: $Q=100$ MeV, dotted curve: $Q=125$ MeV.

\newpage
\begin{figure}[htb]
\vspace*{2cm}
\centerline{\epsfxsize=17cm\epsfbox[5 220 590 620]{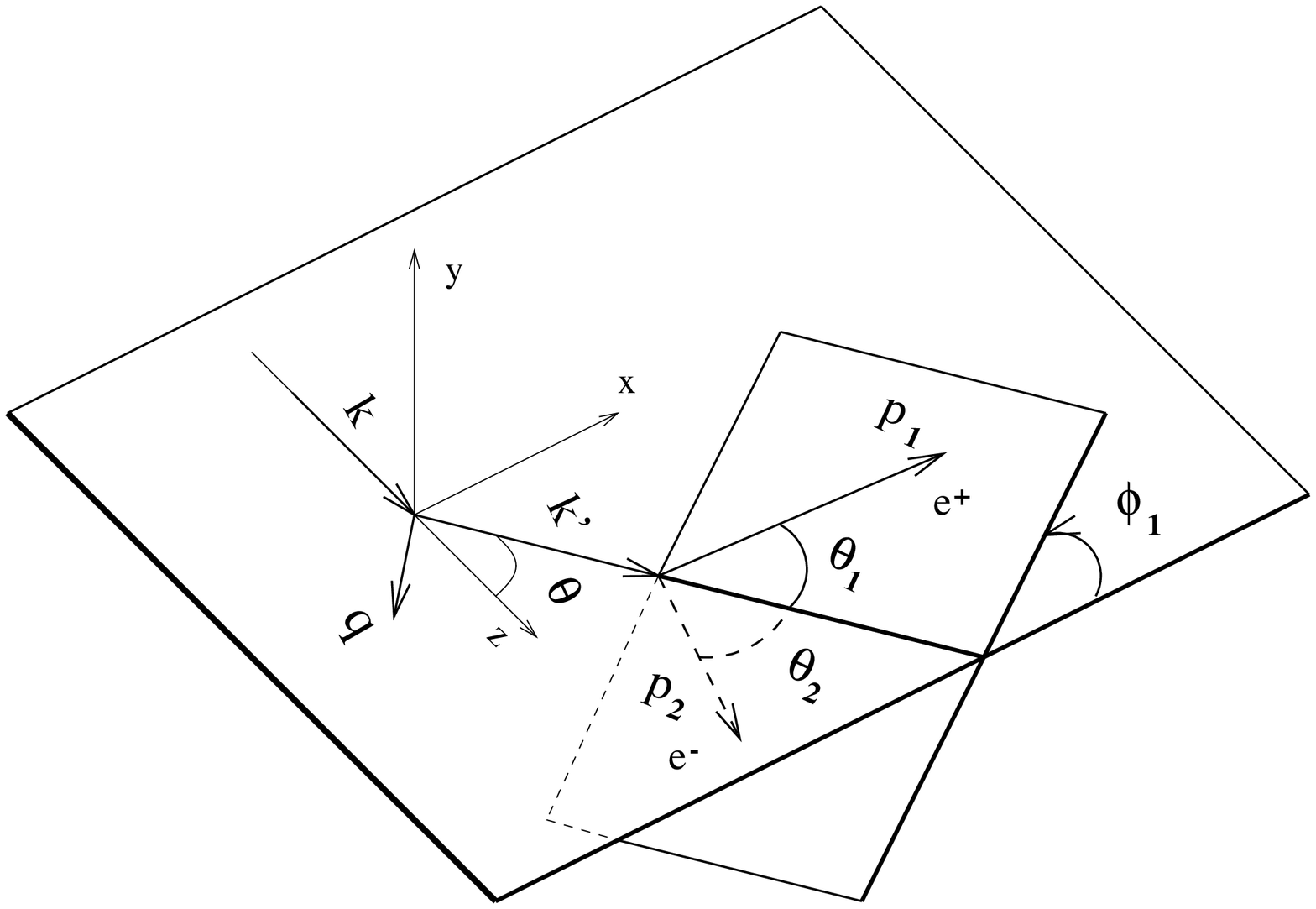}}
\end{figure}
\vspace*{4cm}
\centerline{\bf FIGURE 1}

\newpage
\begin{figure}[htb]
\vspace*{2cm}
\centerline{\epsfxsize=18cm\epsfbox[80 530 520 770]{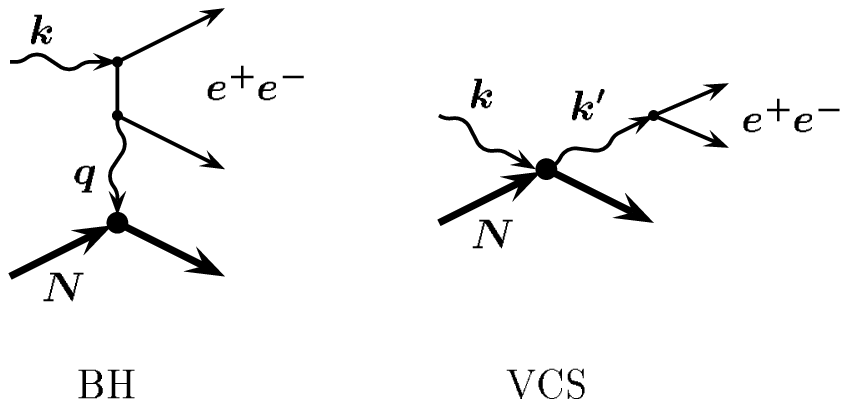}}
\end{figure}
\vspace*{4cm}
\centerline{\bf FIGURE 2}

\newpage
\begin{figure}[htb]
\vspace*{4cm}
\centerline{\epsfxsize=10cm\hspace*{-3cm}\epsfbox{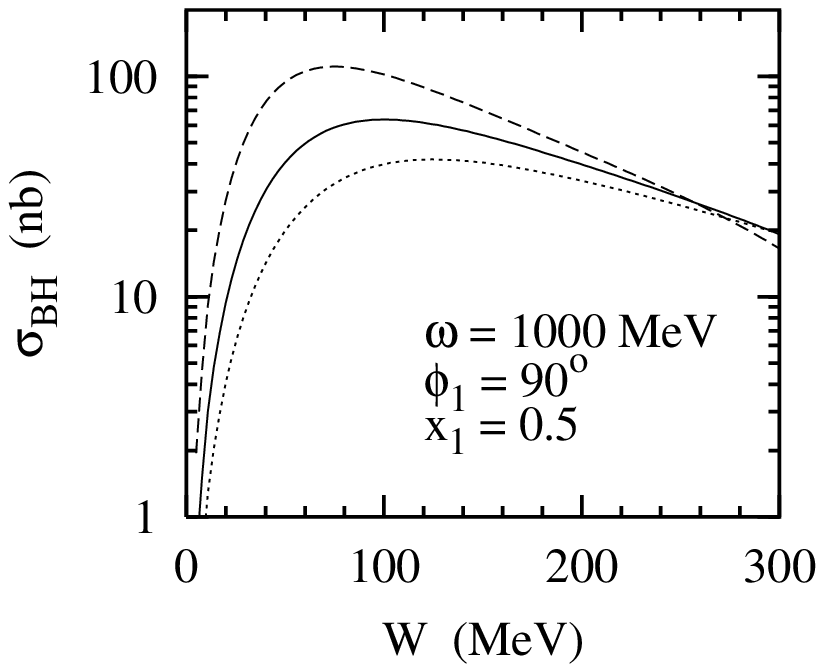}}
\end{figure}
\vspace*{4cm}
\centerline{\bf FIGURE 3}

\newpage
\begin{figure}[htb]
\vspace*{-7cm}
\centerline{\epsfxsize=15cm\hspace*{-3cm}\epsfbox{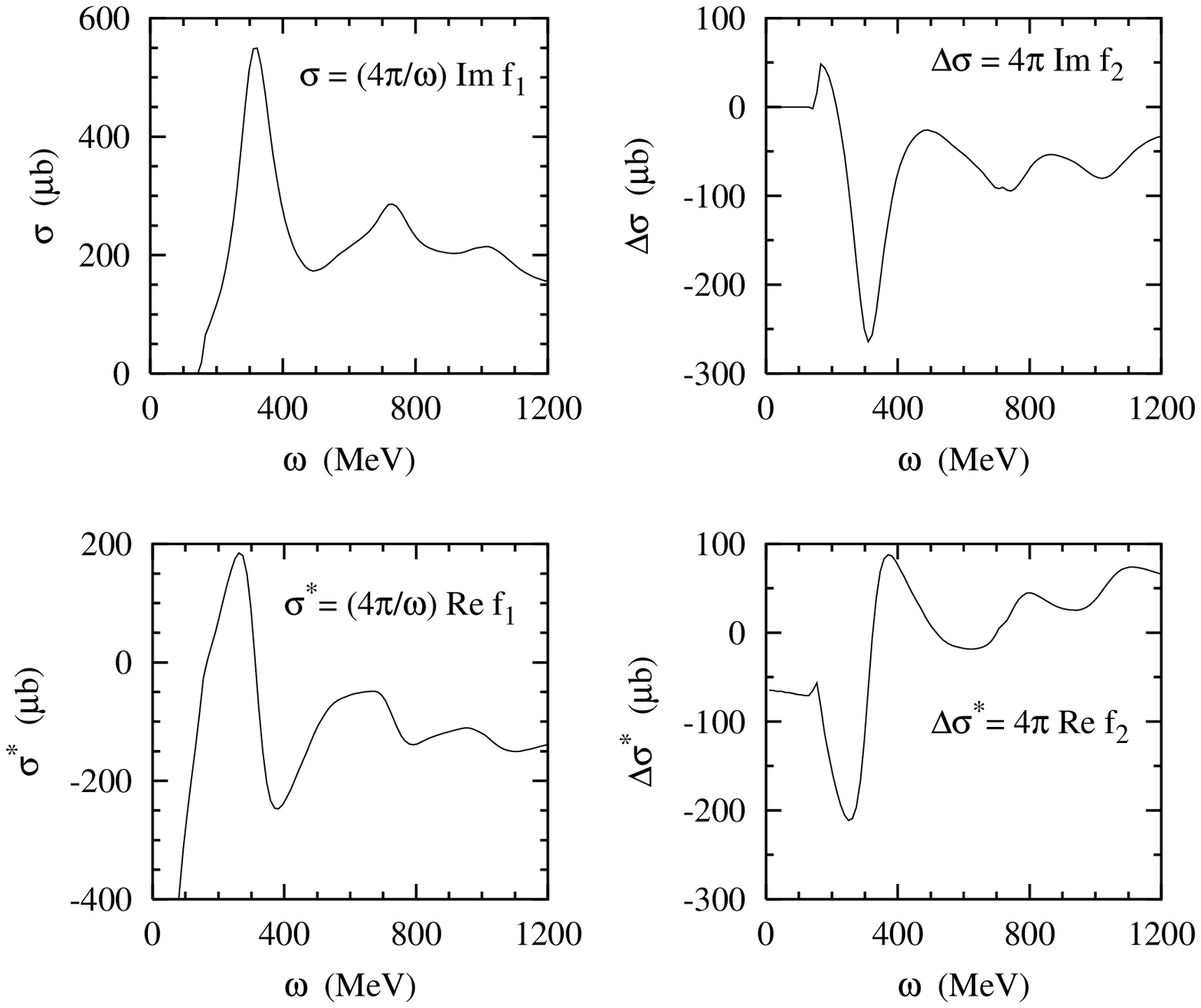}}
\end{figure}
\vspace*{4cm}
\centerline{\bf FIGURE 4}

\newpage
\begin{figure}[htb]
\vspace*{-7cm}
\centerline{\epsfxsize=15cm\hspace*{-3cm}\epsfbox{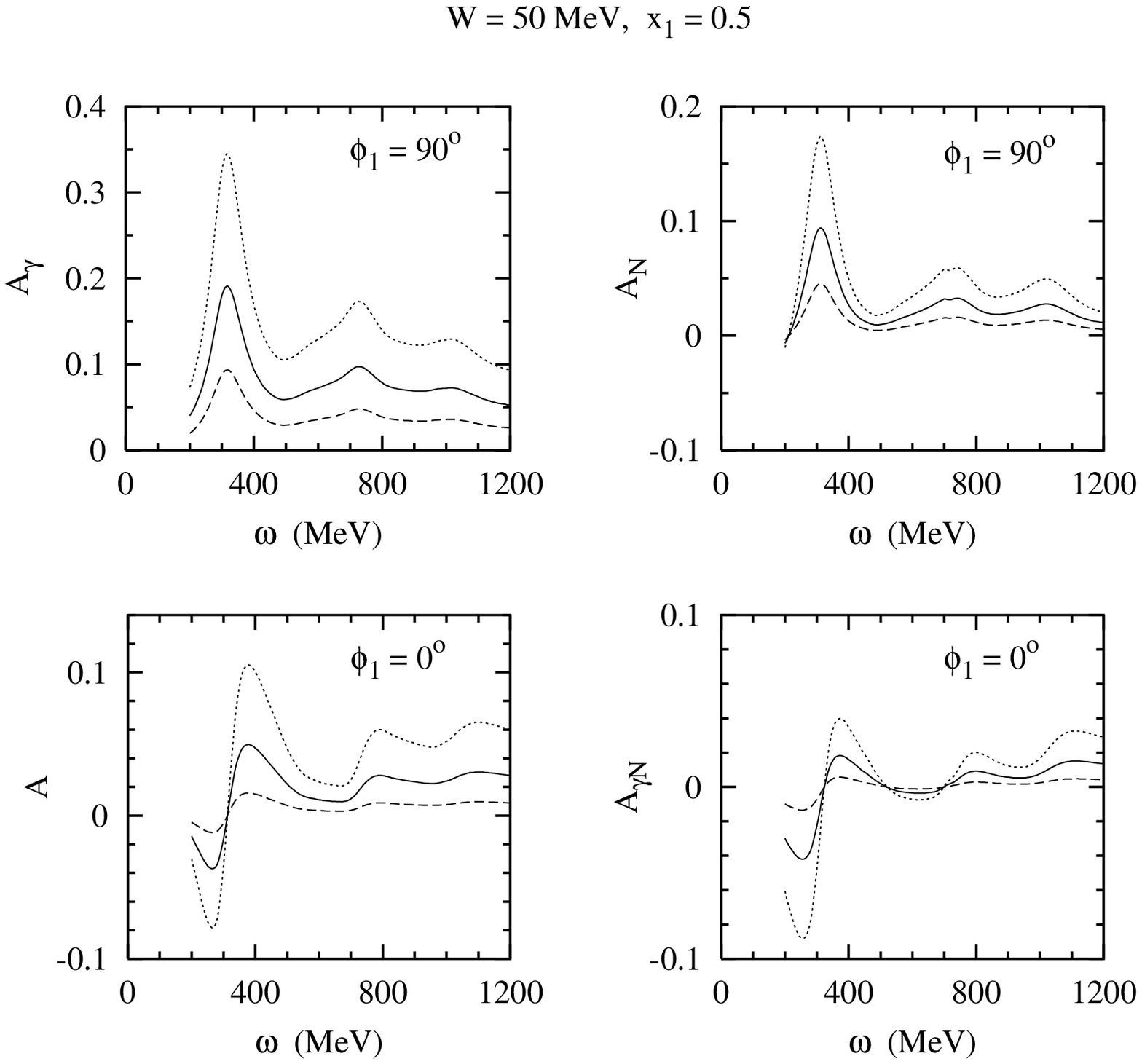}}
\end{figure}
\vspace*{4cm}
\centerline{\bf FIGURE 5}

\end{document}